\documentclass{desyproc}

\usepackage{array}

\begin{document}
\title{On Parton Number Fluctuations}

\author{{\slshape St\'ephane Munier}\\[1ex]
Centre de physique th\'eorique, \'Ecole polytechnique, CNRS, Palaiseau, France}

\contribID{163}

\confID{8648}  
\desyproc{DESY-PROC-2014-04}
\acronym{PANIC14} 
\doi 

\maketitle

\begin{abstract}
Parton evolution with the rapidity essentially is a branching diffusion process. 
We describe the fluctuations of the density of partons which affect the properties of QCD scattering amplitudes at moderately high energies.
We arrive at different functional forms of the latter
in the case of dipole-nucleus and dipole-dipole scattering.
\end{abstract}

\section{Quantum chromodynamics at high density}

Quantum chromodynamics in the high-energy/high-density regime
is a very rich field from a theoretical viewpoint since it involves 
genuinely
nonlinear physics, and nontrivial fluctuations. The
latter are deemed {\em a goldmine
for modern physics} \cite{scapparone}.
From a phenomenological viewpoint, there is a wealth of data from
different experiments which await interpretation.
(For a review, we refer the reader to the recent textbook
by Kovchegov and Levin~\cite{kovchegov2012quantum}).

Electron-proton (or better, nucleus) scattering is
maybe the best experiment to probe QCD in this regime,
as was done at HERA, an $e^\pm p$ facility.
The electron interacts with the proton
 through  a quark-antiquark pair, which appears as a
quantum fluctuation of a (virtual) photon of the Weizs\"acker-Williams field
of the electron.
The probability amplitudes for these fluctuations
follow from a simple QED
calculation. The $q\bar q$ pair is a color dipole, and hence
electron-hadron scattering may be related to dipole-hadron
scattering. If one looks at events in which the $q\bar q$ 
pair has a small-enough size
(as compared to the typical size of a hadron), 
as is possible by selecting longitudinally-polarized
highly-virtual photons,
then perturbative QCD
may be used as a starting point to compute some properties
of the dipole-hadron scattering amplitudes.

As for the interaction of protons and/or nuclei as
is currently performed at the LHC, the observables
need to be carefully chosen if one wants to be able to predict
cross sections from first principles -- at least in the present
state of the art of the theory.
Indeed, one needs a hard momentum scale to justify the use of
perturbation theory, and the latter must be found in the final
state in the form of e.g. the transverse momentum of a jet.
It turns out that an observable such as $p_\perp$-broadening
in proton-nucleus collisions, namely the transverse momentum distribution
of single jets,
may also be related to the dipole-nucleus amplitude.

We will first review the formulation of the rapidity evolution
of the dipole-nucleus scattering amplitude in
QCD in the high-energy limit.
The latter is given by the Balitsky-Kovchegov (BK) equation.
We will relate
the known shape of its solution to gluon-number fluctuations
in the quantum evolution of the dipole.
We will then be able to
predict the form of geometric scaling for dipole-dipole scattering,
which turns out to be different from the solution to the
BK equation.

\section{Dipole-nucleus scattering}

\begin{figure}[h]
\begin{center}
\begin{tabular}{m{0.2\textwidth} | m{0.2\textwidth} m{1cm} m{0.2\textwidth}  }
\includegraphics[width=0.2\textwidth]{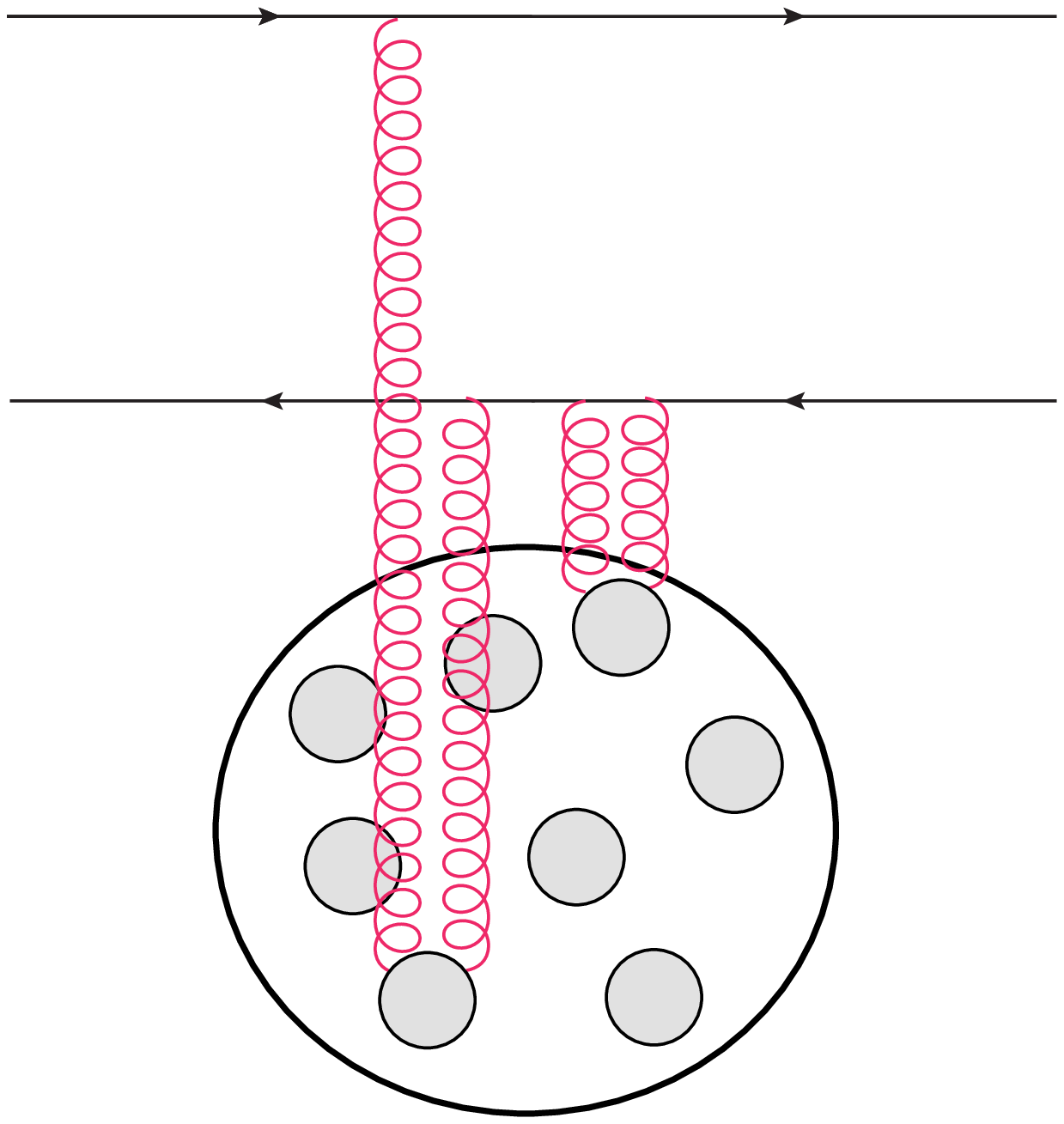} &
\includegraphics[width=0.2\textwidth]{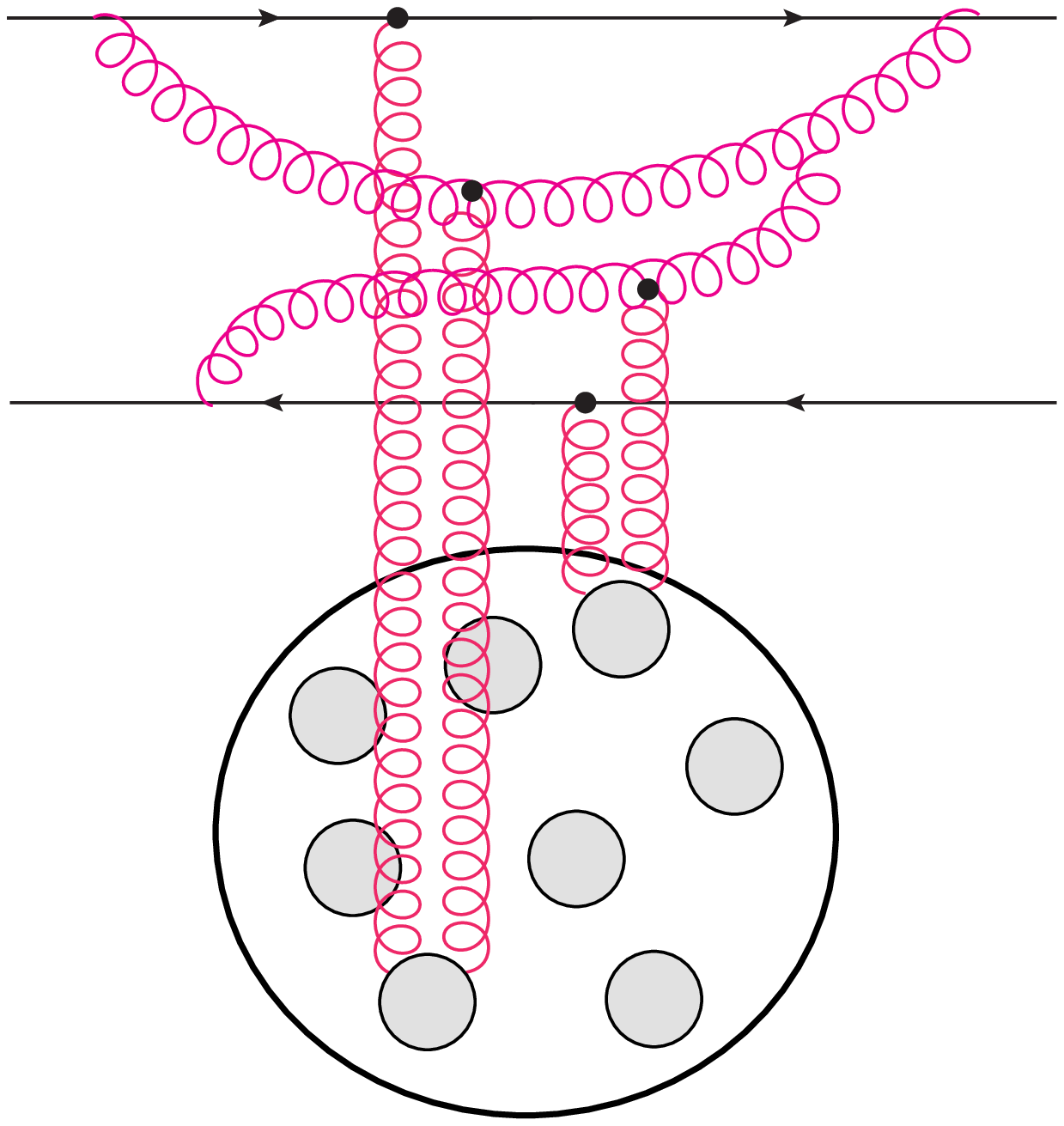}
&$\underset{\begin{minipage}[c]{1.2cm}{\footnotesize large $N_c$}\end{minipage}}{\huge \simeq}$
&\includegraphics[width=0.2\textwidth]{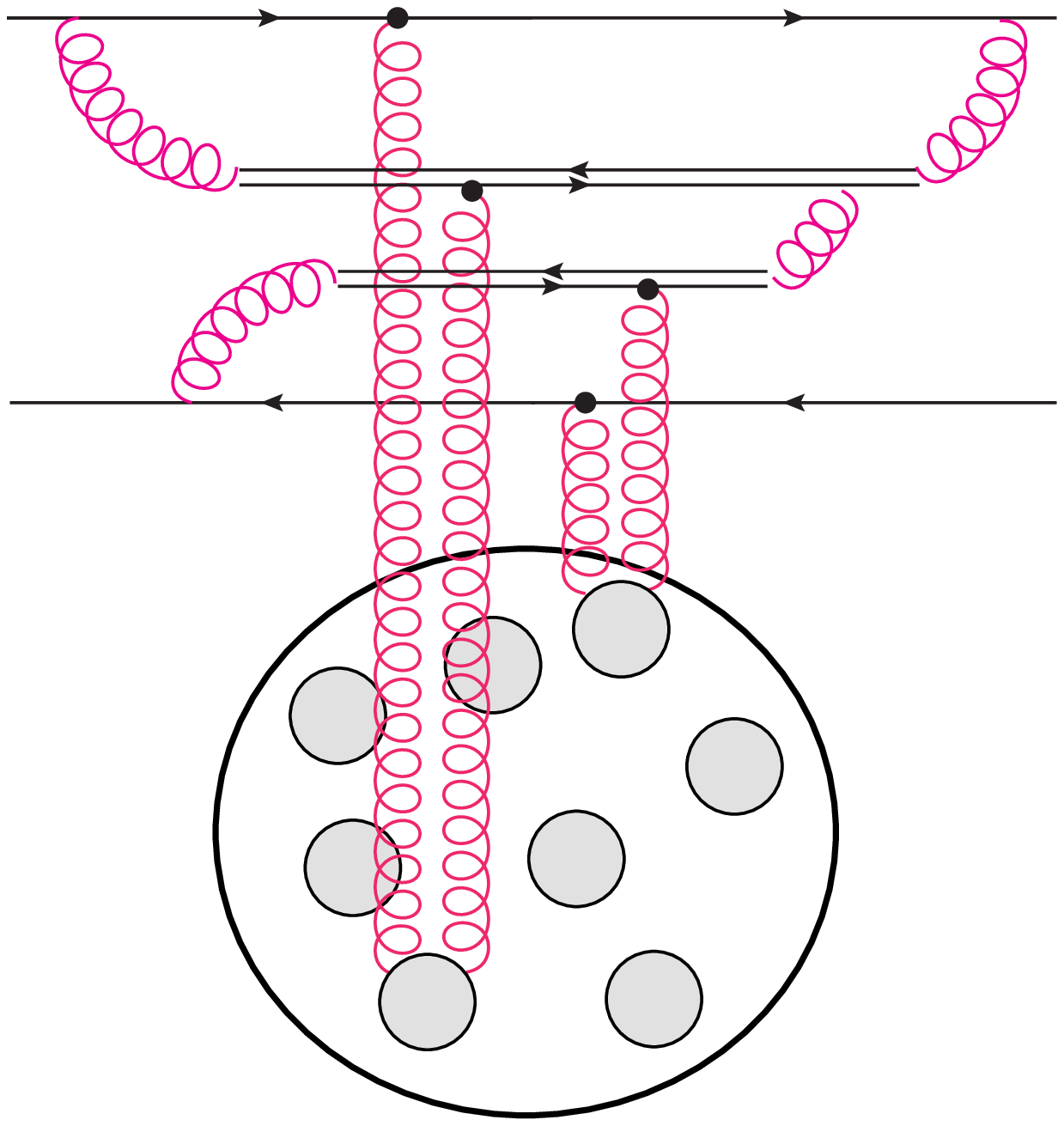}
\\
\\
\multicolumn{1}{c}{(a)} & & \multicolumn{1}{c}{(b)}
\end{tabular}
\end{center}
\caption{Particular graphs contributing to the dipole-nucleus scattering
amplitude at low (a) and high (b) energy in the
restframe of the nucleus.
\label{Fig:dipolenucleus}
}
\end{figure}

Let us start with the scattering of a dipole off a nucleus at relatively low energy. The 
forward elastic amplitude $T$ is a function of the dipole size $r_0$, which is
given by the McLerran-Venugopalan
model:
\begin{equation}
T(r_0)=1-e^{-\frac{r_0^2 Q_A^2}{4}}.
\label{eq:MV}
\end{equation}
This formula resums multiple exchanges of pairs of gluons between 
the bare dipole and the nucleus 
(see Fig.~\ref{Fig:dipolenucleus}a).
$Q_A$ is the saturation momentum of the nucleus.
Equation~(\ref{eq:MV}) essentially means that a dipole of size larger than $\sim 1/Q_A$ is
absorbed ($T\sim 1$), while the nucleus is transparent to dipoles of size smaller 
than $\sim 1/Q_A$.
For our purpose, we may approximate $T(r_0)$ by the step function
$\Theta(\ln r_0^2 Q_A^2/4)$.

Going to higher energies $\sqrt{s}$ 
by increasing the rapidity of the dipole, the scattering process
gets dominated by high-occupancy quantum fluctuations of the initial dipole 
(see Fig.~\ref{Fig:dipolenucleus}b). 
The rapidity ($y\equiv \ln s $) dependence
of the amplitude is given by the Balitsky-Kovchegov (BK) equation
\begin{equation}
\partial_y T(r_0,y)=\bar\alpha\int \frac{d^2 r_1}{2\pi}\frac{r_0^2}{r_1^2(r_0-r_1)^2}
\left[T(r_1,y)+T(r_0-r_1,y)-T(r_0,y)
-T(r_1,y)T(r_0-r_1,y)\right]
\label{eq:BK}
\end{equation}
($\bar\alpha\equiv\frac{\alpha_s N_c}{\pi}$),
whose large-$y$
solutions are traveling waves, namely fronts which translate (almost)
unchanged in shape towards negative values of the $\ln r_0^2$ variable
as the rapidity increases.
The linear part of this equation (the first three terms in the r.h.s.) 
form the BFKL equation, whose kernel
possesses as eigenfunctions the power functions $|r_0|^{2\gamma}$, the corresponding eigenvalues
being $\bar\alpha\chi(\gamma)$, 
where $\chi(\gamma)\equiv 2\psi(1)-\psi(\gamma)-\psi(1-\gamma)$.
Introducing the particular eigenvalue $\chi(\gamma_0)$,
where $\gamma_0$ is such that
$\chi^\prime(\gamma_0)=\chi(\gamma_0)/\gamma_0$, the shape of $T$ as a function 
of the dipole size $r_0$
in the region 
$T\ll 1$ and
the $y$-dependence of the saturation scale
read
\begin{equation}
T(r_0,y)\underset{r_0\ll 1/Q_s(y)}{\sim} 
\ln \frac{1}{r_0^2 Q_s^2(y)}  e^{\gamma_0 \ln ({r_0^2 Q_s^2(y)})}\ \ \text{and}
\ \
Q_s^2(y)/Q_A^2\simeq e^{\bar\alpha\chi^\prime(\gamma_0)y}.
\label{eq:TdA}
\end{equation}

The BK equation~(\ref{eq:BK}) can be established in the framework of the dipole model
(see e.g.~\cite{kovchegov2012quantum}), where gluons
are replaced by zero-size $q\bar q$ pairs. In this model, the
Fock state of the incoming dipole which is ``seen'' by the nucleus at the
time of the interaction is built from successive independent splittings of dipoles.
At a given rapidity $y$, the latter Fock state can be thought of
as a collection of $n$ dipoles, generated by a splitting process which belongs to
a class of processes generically called {\it branching diffusion}.

The main point we wanted to make at this conference and in Ref.~\cite{Mueller:2014fba}
was that $T$ has an
elegant and useful probabilistic
interpretation in the dipole picture: 
{\it It represents the probability that the largest dipole 
present in the Fock state of the incoming $q\bar q$ pair
at the time of the interaction
has a size which is larger than the inverse nuclear saturation momentum,
$1/Q_A$.} 
Indeed, according to the McLerran-Venugopalan model,
a given dipole interacts with the nucleus only if its
size is larger than $1/Q_A$,
hence it is necessary and sufficient that at least 
one of the dipoles in the Fock state
be larger than $1/Q_A$ for the scattering to take place.
Thus solving the BK equation amounts to understanding the statistics of the
extremal particles in a branching random walk (BRW).
Our first task is to recover the shape of the amplitude~(\ref{eq:TdA}),
previously obtained through a analysis
of the BK equation, from the latter statistics.

We observe that the extremal particle in a BRW
has fluctuations which can originate only from two places:
From the first stages of the rapidity evolution,
when the overall number of dipoles is small and
thus subject to large statistical fluctuations (we shall call this type
of stochasticity
``front fluctuations''), and from the tip of the distribution, where
by definition, particle numbers keep small.
Elsewhere, the evolution is essentially 
deterministic since it acts on a large number
of objects.
\begin{figure}[h]
\begin{center}
\begin{tabular}{m{0.45\textwidth} m{0.55\textwidth}}
\includegraphics[width=0.4\textwidth]{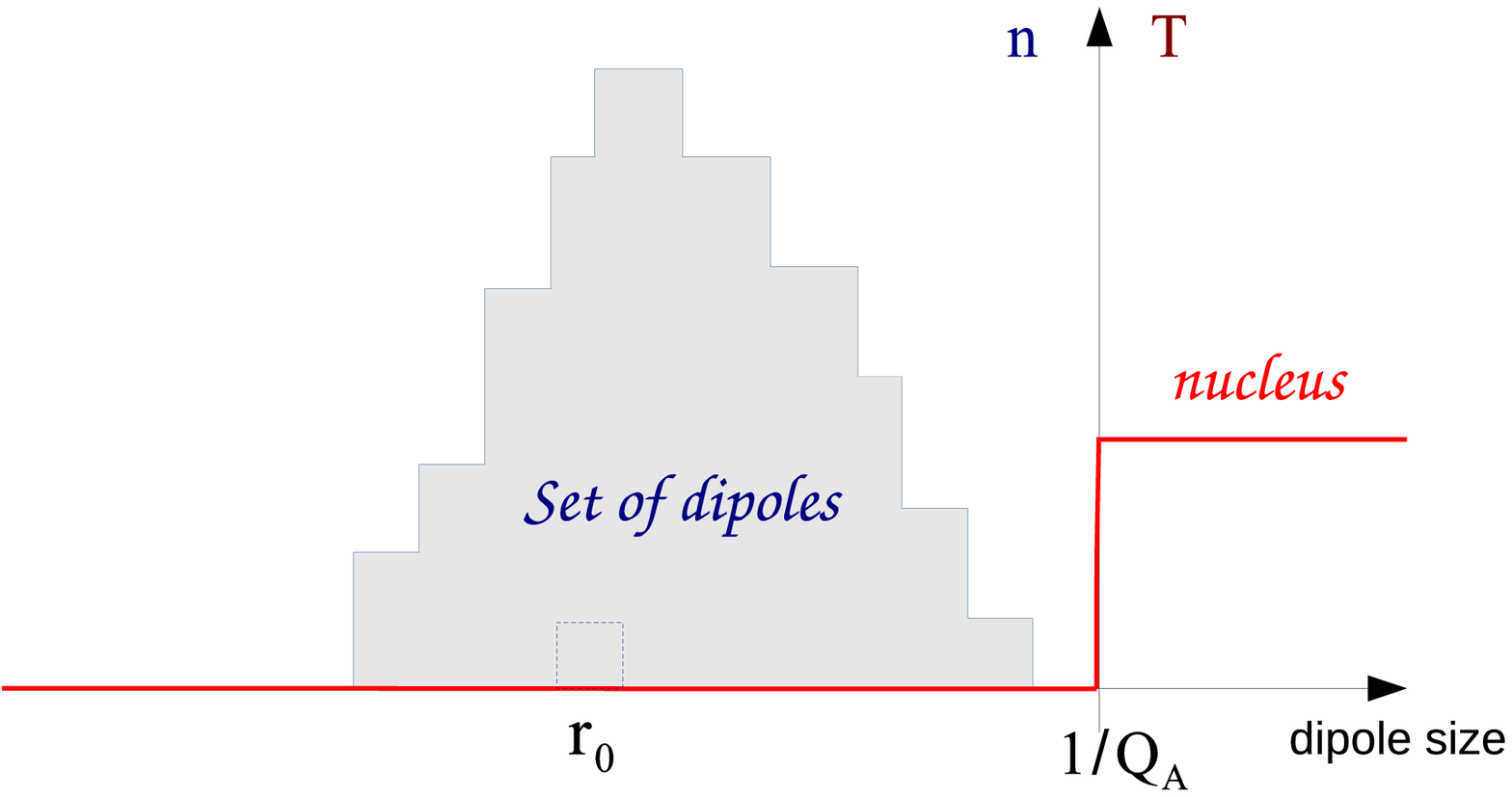} &
\includegraphics[width=0.5\textwidth]{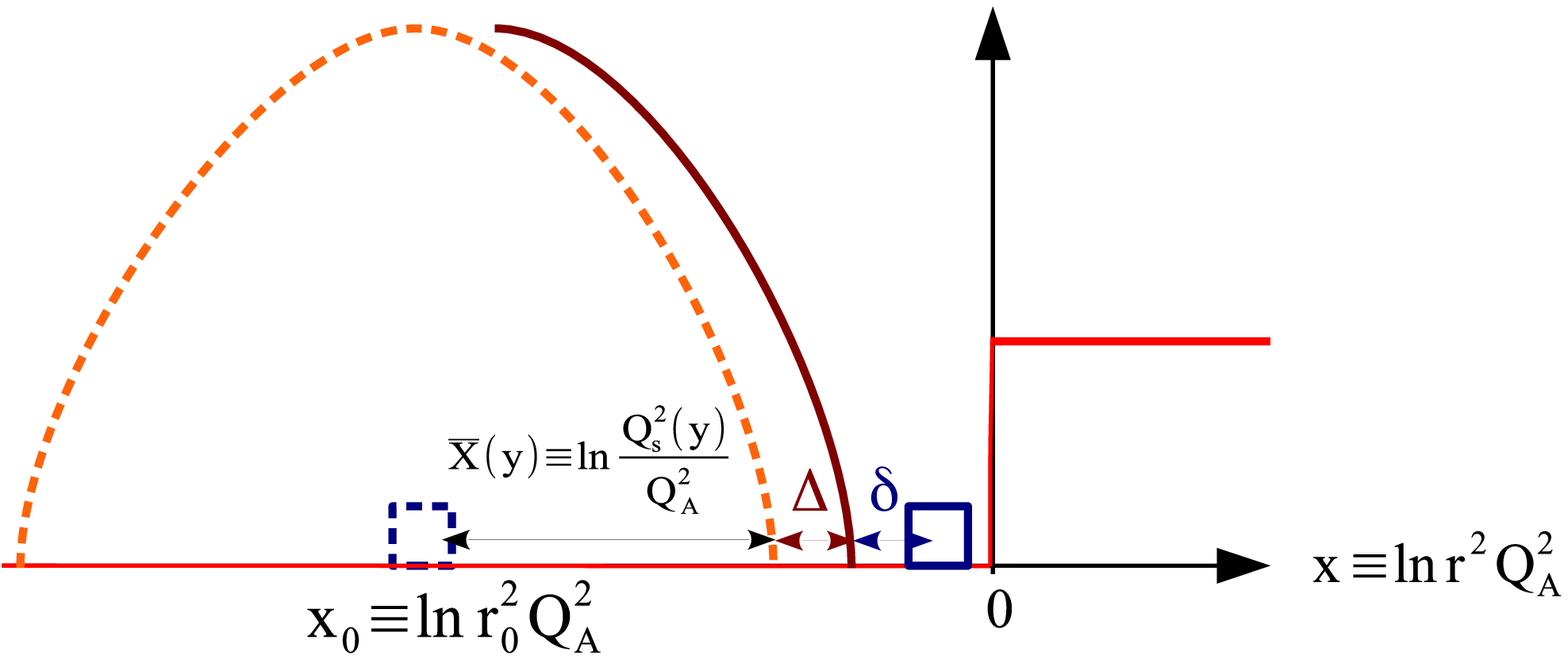}\\
\multicolumn{1}{c}{(a)} & \multicolumn{1}{c}{(b)}
\end{tabular}
\end{center}
\caption{\label{Fig:d1}
(a) Sketch
of the
dipole distribution (as seen at a given impact parameter) 
in a particular realization of the evolution to the rapidity $y$
of an initial dipole of size $r_0$.
(b) Model for the evolution of a given realization, with the ``front'' and ``tip'' 
fluctuations described in the text.
}
\end{figure}
The effect of the front fluctuations is to shift the particle distribution by~$\Delta$.
We conjecture\footnote{Arguments in favor of this conjecture 
were presented in Ref.~\cite{Mueller:2014gpa}.} 
that the distribution of $\Delta$ is $\propto e^{-\gamma_0\Delta}$.
The effect of the tip fluctuations is instead
to send randomly particles ahead of the front
by $\delta$. We conjecture the same exponential law $\propto e^{-\gamma_0\delta}$.

We introduce our notations
in Fig.~\ref{Fig:d1}. According to the previous discussion, 
in a particular event, the scattering occurs
if $x_0+\bar X(y)+\Delta+\delta\geq 0$.
Hence the amplitude $T$ simply is the average of this condition over
$\Delta$ and $\delta$:
\begin{equation}
T\propto \int_{0}^{+\infty}
d\delta\,e^{-\gamma_0\delta}
\int_{0}^{+\infty}
d\Delta\,e^{-\gamma_0\Delta}
\Theta(x_0+\bar X(y)+\Delta+\delta)\propto(-x_0-\bar X(y))e^{\gamma_0(x_0+\bar X(y))}.
\end{equation}
Switching back to the QCD variables, we recover the expression of $T$ given in
Eq.~(\ref{eq:TdA}).
We conclude that
{\it the shape of the dipole-nucleus scattering amplitude 
as a function of the dipole size is 
directly related to the event-by-event fluctuations of the
size of the largest dipole}, which in turn stem from
the fluctuations of the numbers of gluons produced in the QCD evolution.

\section{Dipole-dipole scattering}

\begin{figure}[h]
\begin{center}
\begin{tabular}{m{0.22\textwidth} m{0.35\textwidth} m{0.35\textwidth}}
\includegraphics[width=0.2\textwidth]{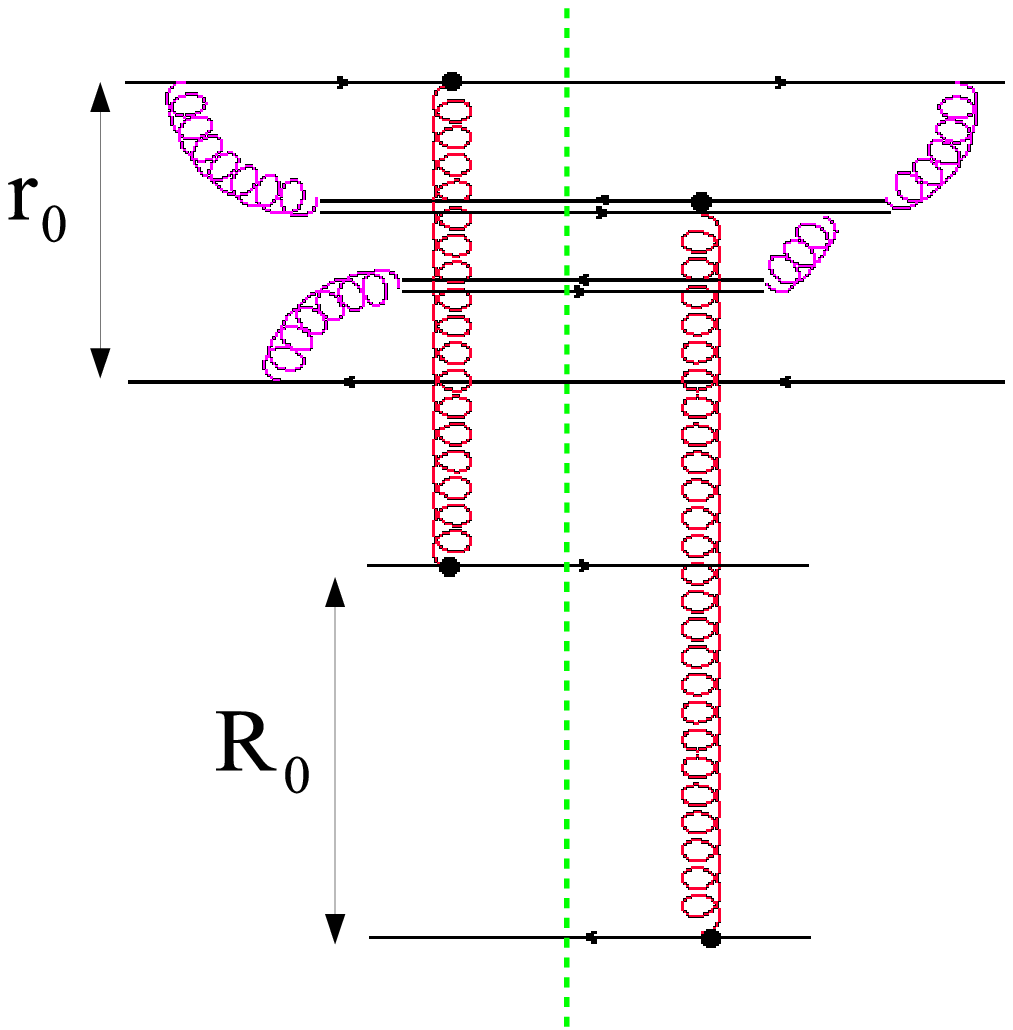} &
\includegraphics[width=0.32\textwidth]{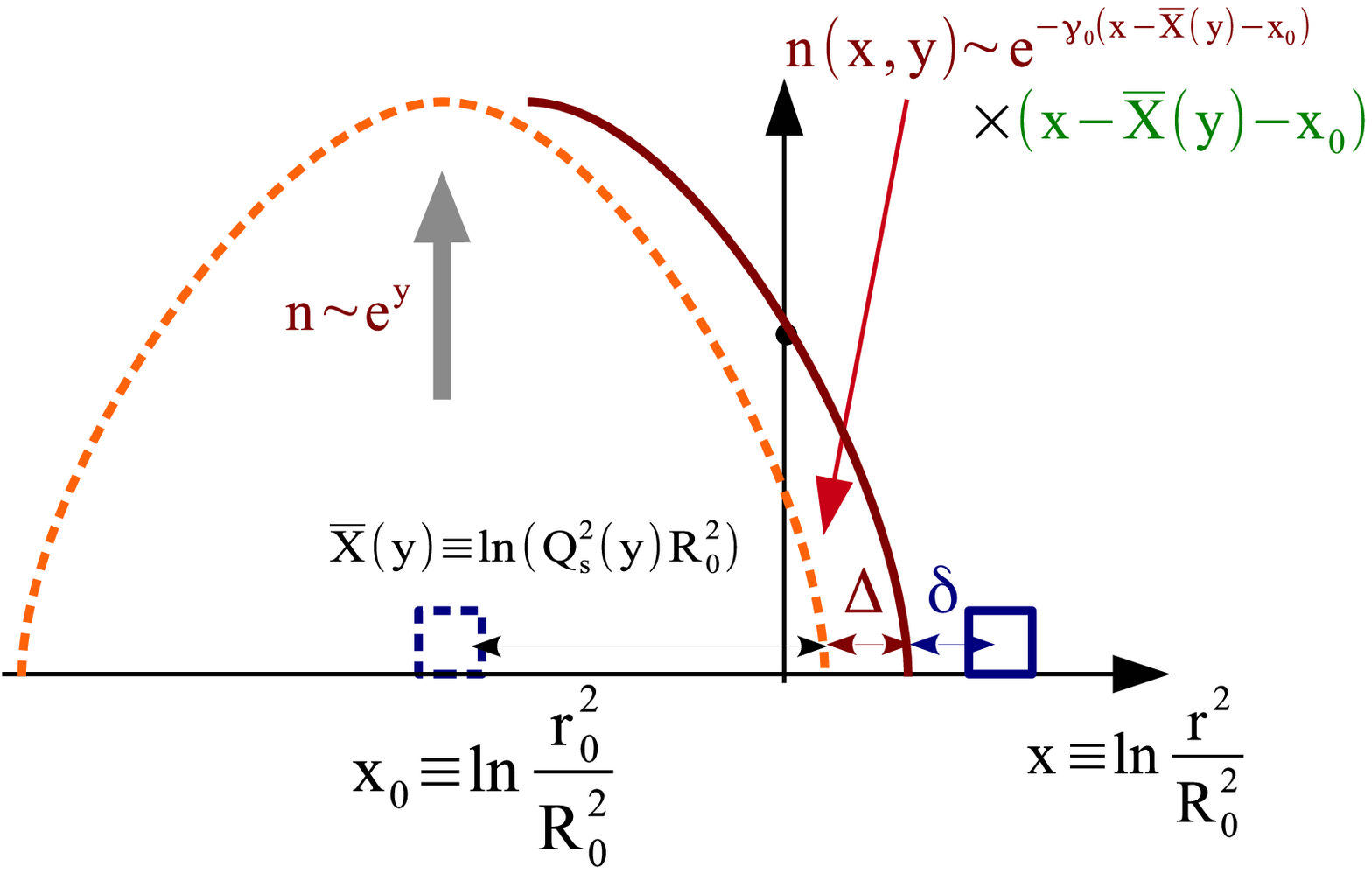} &
\includegraphics[width=0.32\textwidth]{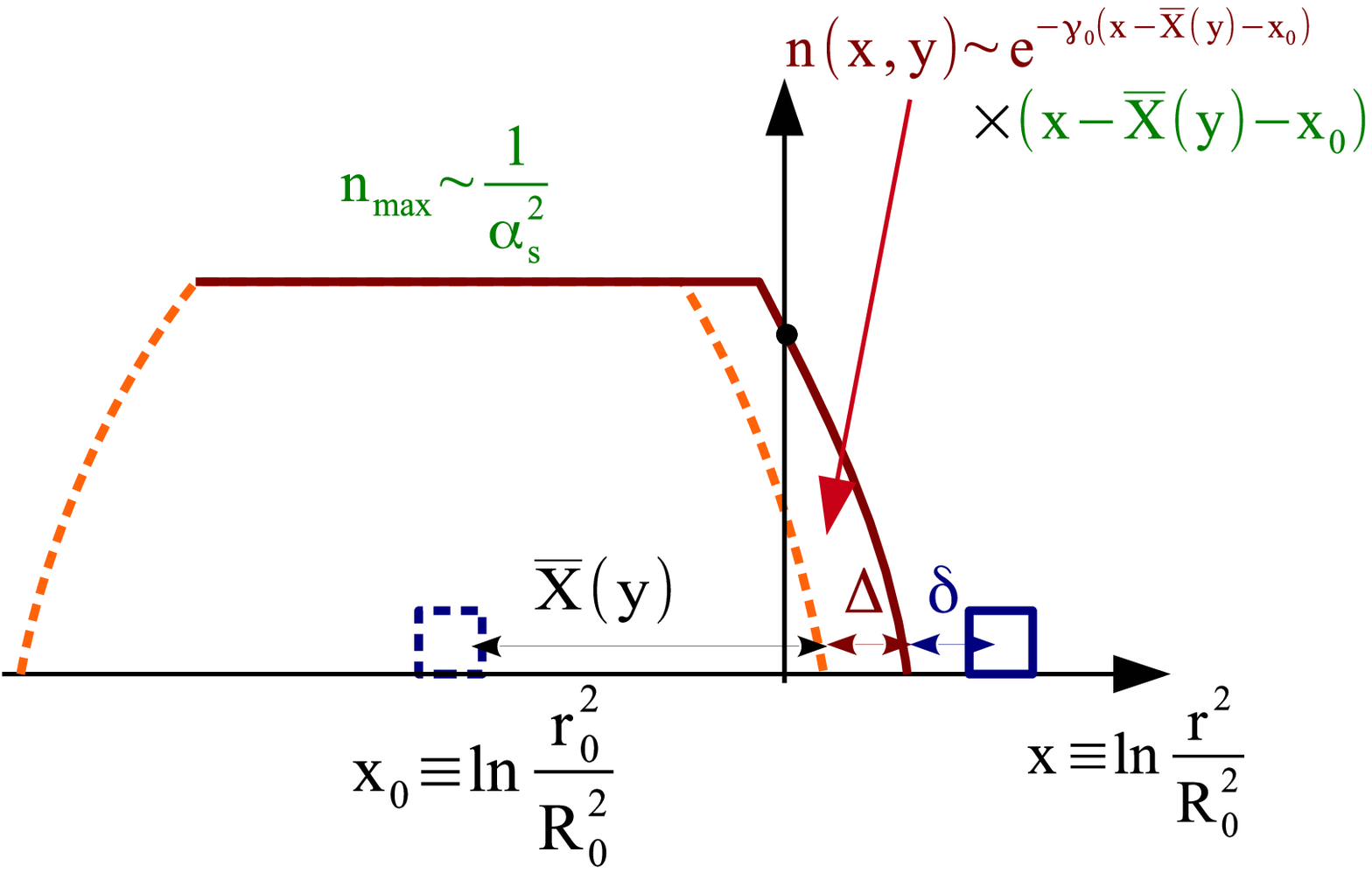}\\
\multicolumn{1}{c}{(a)} & \multicolumn{1}{c}{(b)} & \multicolumn{1}{c}{(c)}
\end{tabular}
\end{center}
\caption{\label{Fig:dd} (a) Graph contributing to dipole-dipole scattering
at high energies. (b) Sketch of the evolution of the dipole number density, model including
fluctuations. (c) The same, but with saturation.
}
\end{figure}

While the dipole-nucleus amplitude probes the statistics of the largest dipole in the
quantum evolution, the physics of dipole-dipole scattering is a bit different: 
Indeed, since the elementary  amplitude (for dipoles of respective sizes $r_0$ and $R_0$)
at zero rapidity is essentially
$T(r_0,R_0)\sim \alpha_s^2\delta(\ln r_0^2/R_0^2)$, 
it is the very shape of the dipole number distribution
that is actually probed (Fig.~\ref{Fig:dd}). So in order to compute the shape of the amplitude,
we need on one hand the probability distribution of the front fluctuations used before, and 
on the other hand the
shape of the dipole number density from the deterministic evolution.
We also need to implement {\it saturation} in the evolution (see Fig.~\ref{Fig:dd}c)
to comply with the unitarity constraint
$T\leq 1$. All in all, we obtain
\begin{equation}
T(r_0,y)\underset{r_0\ll 1/Q_s(y)}{\sim} 
\ln^2 \frac{1}{r_0^2 Q_s^2(y)}  e^{\gamma_0 \ln ({r_0^2 Q_s^2(y)})}
\ \ \text{where}
\ \
Q_s^2(y)R_0^2\simeq e^{\bar\alpha\chi^\prime(\gamma_0)y}.
\label{eq:Tdd}
\end{equation}
Interestingly enough, it differs from the dipole-nucleus case;
compare Eq.~(\ref{eq:TdA}) to Eq.~(\ref{eq:Tdd}).
This is the main prediction of the way of looking at QCD evolution we have promoted at this
conference and in Ref.~\cite{Mueller:2014fba}.

We refer the reader to~\cite{Mueller:2014fba} for the details, references,
and more results, in particular on the finite-$y$ corrections to the saturation scale
in both the dipole-dipole and dipole-nucleus cases.


\begin{footnotesize}


\end{footnotesize}


\end{document}